\documentclass[preprint,showpacs,preprintnumbers,amsmath,amssymb,aps,nofootinbib]{revtex4}
\usepackage[latin9]{inputenc}
\usepackage{color}
\usepackage{esint}
\PassOptionsToPackage{normalem}{ulem}
\usepackage{ulem}
\usepackage[unicode=true,
 bookmarks=true,bookmarksnumbered=false,bookmarksopen=false,
 breaklinks=false,pdfborder={0 0 0},backref=false,colorlinks=true]
 {hyperref}
\hypersetup{
 linkcolor=blue, citecolor=red}
\usepackage{breakurl}

\makeatletter
\@ifundefined{textcolor}{}
{%
 \definecolor{BLACK}{gray}{0}
 \definecolor{WHITE}{gray}{1}
 \definecolor{RED}{rgb}{1,0,0}
 \definecolor{GREEN}{rgb}{0,1,0}
 \definecolor{BLUE}{rgb}{0,0,1}
 \definecolor{CYAN}{cmyk}{1,0,0,0}
 \definecolor{MAGENTA}{cmyk}{0,1,0,0}
 \definecolor{YELLOW}{cmyk}{0,0,1,0}
}

\usepackage{dcolumn}
\usepackage{bm}

\usepackage[english]{babel}

\makeatother

\begin{document}

\title{Relativistic viscous hydrodynamics order by order}

\author{Jian-Hua Gao}

\email{gaojh@sdu.edu.cn}

\selectlanguage{english}%

\affiliation{Shandong Provincial Key Laboratory of Optical Astronomy and Solar-Terrestrial
Environment,  Institute of Space Sciences, Shandong University, Weihai 264209, China}

\affiliation{Key Laboratory of Quark and Lepton Physics (MOE), Central China Normal
University, Wuhan 430079, China}

\author{Shi Pu}
\affiliation{
Institute for Theoretical Physics, Goethe University,
Max-von-Laue-Stra{\ss}e 1, 60438 Frankfurt am Main, Germany}

\affiliation{Department of Physics, National Center for Theoretical Sciences,
and Leung Center for Cosmology and Particle Astrophysics, National
Taiwan University, Taipei 10617, Taiwan}

\affiliation{Interdisciplinary Center for Theoretical Study and Department of
Modern Physics, University of Science and Technology of China, Hefei
230026, China}
\begin{abstract}
In this paper, we propose a method of solving the viscous hydrodynamics
order by order in a derivative expansion. In such a method, the zero-order solution is just  one of the ideal hydrodynamics. All the
other higher order corrections satisfy the same first-order partial
differential equations but with different
inhomogeneous terms.  We take the Bjorken flow as an example to test the validity of our method  and
present how to deal with  the problems about  the initial condition and perturbation evolution in our formalism.
\end{abstract}


\maketitle

\section{Introduction}

Relativistic hydrodynamics has been an important and useful theoretical
tool in high energy heavy-ion physics such as at BNL Relativistic Heavy Ion Collider (RHIC) and CERN Large Hadron Collider (LHC), which
have succeeded greatly in describing the collective flow from the
data of those colliders \cite{Heinz:2013th,Gale:2013da,Teaney:2009qa,Romatschke:2007mq,Song:2007fn,Dusling:2007gi,Molnar:2008xj,Schenke:2010rr}.
Hydrodynamics can be considered as an macroscopic effective field
theory of more fundamental microscopic theory such as quantum field
theory, in describing the non-equilibrium evolution of a given system.
However, it is not {trivial}
to build a consistent and causal relativistic hydrodynamics beyond
the ideal hydrodynamics. The relativistic first order viscous hydrodynamics
was first proposed by Eckart in \cite{Eckart:1940te}, and Landau
and Lifshitz in \cite{Landau:book}, both of which showed that dissipative
fluctuation may propagate at an infinite speed, which is inconsistent
with the relativistics. This is the so-called causality problem. In addition,
the solution is also unstable due to small perturbation of the equilibria
in these viscous hydrodynamics \cite{Hiscock:1983zz}. The nonrelativistic
causal viscous hydrodynamics was first presented by Müller in \cite{Muller:1967}
and was later generalized into the relativistic version by Israel
and Stewart in \cite{Israel:1979wp}. They remedied the previous viscous
hydrodynamics by introducing some second order terms in deviations
away from equilibrium into the entropy current. Therefore these formalisms
are also referred to as the second order theory of viscous hydrodynamics.
{For example, the relaxation time for shear viscous
tensor, which is one of the well-known second order parameters, describes
how long it will take for the system to return to the equilibrium states after a
small perturbation via shear viscosity. Therefore, if all these hydrodynamic
parameters satisfy certain constraints, the system will be causal and
stable (e.g., for shear viscous tensor \cite{Pu:2009fj,Pu:2011vr},
bulk viscous pressure \cite{Denicol:2008ha}, and heat conducting flow
\cite{Pu:2010zz,Pu:2011vr}).} Recently, some other authors  discussed
further the second order viscous hydrodynamics \cite{Baier:2007ix,Betz:2008me,Betz:2009zz},
especially from the point of view of effective theory, in which all
the second derivative terms are included.

There exist two different methods solving
the viscous hydrodynamic equations, either expressing the dissipative
contributions in energy-momentum tensor or charge current in terms
of differentials on primary variables such as fluid $4$-velocity
$u^{\mu}$ and chemical potential, then substituting them into the
hydrodynamic equation, or regarding the dissipative quantities as
independent dynamical quantities which satisfies extra differential
equations. {The same difficulty in both methods
is that the contributions from different orders are mixed together,
which implies small errors in high orders might also cause big uncertainty
in the numerical simulations after time evolutions. On the other hand,
the point of view of effective theory, where higher order terms should
always be small corrections to the lower order during the whole evolution,
might give us some hints to simplify these problem. Besides, from
the second order to the third  or even higher orders, one has
to deal with more and more complicated differential hydrodynamic equations.}

In {Sec. II} of this paper, we will try to present
a consistent formalism of solving the viscous hydrodynamic equation
order by order {in  comparison with microscopic
theories}. We will show that the zero order solution is just the one
of the ideal hydrodynamics in our method and all the other higher
order corrections satisfy the same first-order partial differential
equation but with different {inhomogeneous} source
terms. We find that our method is a recursion process, the next order
solution can be obtained only after we get all the previous order
solution. In every order calculation, we only need to deal with the
same first order differential equations with different
{inhomogeneous} source terms. Such method can be manipulated
to any higher order.
In Sec.III, we will discuss how to deal with the problems about the initial condition  and stability in our formalism.
In Sec.IV, we choose the Bjorken flow as a test to illustrate the validity of our method and how to manipulate
 the initial condition  and perturbation evolution  specifically. Finally, there is  the conclusion in Sec V.

\section{Hydrodynamics order by order}

Since {we will present our method}
mainly theoretically or formally,
{for simplicity,} we will restrict ourselves to the
conformal non-charged fluid. In such a system, the dissipative terms
are constrained greatly due to the conformal symmetry. {More
general cases can be extended straightforwardly and will be presented
elsewhere.} Since the fluid is not charged, only energy-momentum conservation
is involved,
\begin{eqnarray}
\partial_{\nu}T^{\nu\mu} & = & 0,\label{div-Tmn}
\end{eqnarray}
where {the energy-momentum tensor} $T^{\mu\nu}$ is
assumed to be able to expand as the primary hydrodynamic variable,
local fluid velocity $u^{\mu}(x)$ ($u^{2}=-1$) and local temperature
$T(x)$. In the following, we will always work in the Landau frame
and adopt the convention of the metric tensor $g^{\mu\nu}=[-1,+1,+1,+1]$.
In such frame and convention, the energy-momentum tensor can be generally
decomposed into
\begin{eqnarray}
T^{\mu\nu} & = & \left(\epsilon+P\right)u^{\mu}u^{\nu}+Pg^{\mu\nu}+\Pi^{\mu\nu},\label{Tmunu}
\end{eqnarray}
where {$\epsilon$ is the energy density, $P$ is
the pressure, and} $\Pi^{\mu\nu}$ includes all the dissipative terms
and satisfies $u_{\mu}\Pi^{\mu\nu}=0$.

{Generally, in long wavelength and low frequency limit,
if Knudsen number $K=\ell_{mfp}\partial_{\mu}\ll1$, with $\ell_{mfp}$
the mean free path of particles and $\partial_{\mu}$ the space-time
derivatives, the hydrodynamic is workable \cite{Betz:2008me,Betz:2009zz}.
In this case, we can expand all hydrodynamic quantities and equations
in the power of the Knudsen number. In the leading order, we will get
the ideal fluid. In the first order, $\Pi^{\mu\nu}$ will be introduced
and can} be expanded as the differentials of the local velocity $u^{\mu}$
order by order. In a conformal theory, this dissipative term can be
generally written as \cite{Baier:2007ix,York:2008rr},

\begin{eqnarray}
\Pi^{\mu\nu} & = & -\eta\sigma^{\mu\nu}+\pi^{\mu\nu},\label{Pimunu-1}\\
\sigma^{\mu\nu} & \equiv & \Delta^{\mu\alpha}\Delta^{\nu\beta}\left(\partial_{\alpha}u_{\beta}+\partial_{\beta}u_{\alpha}-\frac{2}{3}\Delta_{\alpha\beta}\nabla\cdot u\right),\\
\Delta^{\mu\nu} & \equiv & g^{\mu\nu}+u^{\mu}u^{\nu},\ \ \nabla\cdot u\equiv\Delta^{\alpha\beta}\partial_{\alpha}u_{\beta},
\end{eqnarray}
where $\eta$ is the shear viscosity, $\pi^{\mu\nu}$ is the second-order
differential terms and in a conformal theory, can be generally decomposed
into the following form,
\begin{eqnarray}
\pi^{\mu\nu} & = & \eta\tau_{\Pi}\left[u^{\alpha}\partial_{\alpha}\sigma^{\mu\nu}+\frac{1}{3}\sigma^{\mu\nu}\partial_{\alpha}u^{\alpha}\right]+\lambda_{1}\left[{\sigma^{\mu}}_{\alpha}\sigma^{\nu\alpha}-\frac{1}{3}\Delta^{\mu\nu}\sigma_{\alpha\beta}\sigma^{\alpha\beta}\right]\nonumber \\
 &  & +\frac{1}{2}\lambda_{2}\left[{\sigma^{\mu}}_{\alpha}\Omega^{\nu\alpha}{\sigma^{\nu}}_{\alpha}\Omega^{\mu\alpha}\frac{}{}\right]+\lambda_{3}\left[{\Omega^{\mu}}_{\alpha}\Omega^{\nu\alpha}-\frac{1}{3}\Delta^{\mu\nu}\Omega_{\alpha\beta}\Omega^{\alpha\beta}\right],\\
\Omega_{\mu\nu} & \equiv & \frac{1}{2}\Delta_{\mu\alpha}\Delta_{\nu\beta}\left(\partial^{\alpha}u^{\beta}-\partial^{\beta}u^{\alpha}\right),
\end{eqnarray}
where {$\tau_{\Pi},\lambda_{1,2,3}$ are transport
coefficients in the second order theory and $\Omega^{\mu\nu}$ is
the vorticity tensor.} The entropy current $S^{\mu}$ is defined as
{\cite{Israel:1979wp}} {
\begin{eqnarray}
S^{\mu} & = & \frac{P}{T}u^{\mu}-\frac{1}{T}u_{\nu}T^{\nu\mu}-Q^{\mu},\label{Smu}
\end{eqnarray}
}where $Q^{\mu}$ represents a possible second order correction. {In
the leading order, $S^{\mu}=su^{\mu}$, with $s$ the entropy density
\begin{eqnarray}
sT=\epsilon+P.\label{s}
\end{eqnarray}
}

{Now we will  propose our method. Firstly,
it is quite natural and straightforward that} we will treat fluid
$4$-velocity $u^{\mu}(x)$ and temperature $T(x)$ as the primary
variables, energy density and pressure can be expressed as the function
of $T(x)$ by the equation of state. We can imagine the final solution
of $u^{\mu}(x)$ and $T(x)$ can be obtained by the serials expansion
as
\begin{eqnarray}
u^{\mu}(x) & = & u_{0}^{\mu}(x)+u_{1}^{\mu}(x)+u_{2}^{\mu}(x)+...,\\
T(x) & = & T_{0}(x)+T_{1}(x)+T_{2}(x)+...,
\end{eqnarray}
where the series are expanded in the power of Knudsen number. {Then, we assume that
all hydrodynamic quantities and equations can be expanded in the power
of Knudsen number. For example,} it follows that the energy-momentum
tensor can expanded as
\begin{eqnarray}
T^{\nu\mu}=T_{0}^{\nu\mu}+T_{1}^{\nu\mu}+T_{2}^{\nu\mu}+...\label{Tmunu-series}
\end{eqnarray}
The zero-order energy-momentum tensor is given by
\begin{eqnarray}
T_{0}^{\nu\mu}=\left(\epsilon_{0}+P_{0}\right)u_{0}^{\mu}u_{0}^{\nu}+P_{0}g^{\mu\nu}\label{Tmunu-0}
\end{eqnarray}
where $\epsilon_{0}\equiv\epsilon(T_{0})$ and $P_{0}\equiv P(T_{0})$.
It is just the decomposition of the ideal fluid.

{Secondly, in order to avoid the mixture of different
orders, we assume the differential hydrodynamic equations satisfy
the conservation law order by order, i.e. we let
\begin{equation}
\partial_{\mu}T_{i}^{\mu\nu}=0, \ \ \ (i=0,1,2,...).
\end{equation}
It looks very robust and adds more constraints to the hydrodynamic
equations, but it is reasonable. From the classical kinetic theory,
i.e., the Boltzmann equations, the distribution function $f$ can be
expanded in power of $K$, $f=f_{0}+f_{1}+f_{2}+...$ and obtained
order by order. Provided the time reversal symmetry is protected,
we can get $\partial_{\mu}T_{i}^{\mu\nu}=0$ (also see Appendix \ref{sec:Order-expansion-in}).
These kinds of methods are widely used in theoretical physics; e.g.,
for quantum kinetic theory, a similar treatment will give the exact
transport coefficients of chiral magnetic and vortical effects \cite{Gao:2012ix,Chen:2012ca}
or Hall effects \cite{Chen:2013dca}, and other related hydrodynamics
\cite{Bhattacharyya:2008jc,Banerjee:2008th,Floerchinger:2014hua}.) }

{Back to our case, }the zero-order approximation $u_{0}^{\mu}$
and $T_{0}$ can be obtained by solving the ideal hydrodynamic equation,
\begin{eqnarray}
\partial_{\nu}T_{0}^{\nu\mu}=0.\label{equation-0}
\end{eqnarray}
As usual, we can decompose them into a component parallel to $u_{0}^{\mu}$
by contracting Eq.(\ref{equation-0}) with $u_{0}^{\mu}$,
\begin{eqnarray}
\left(\epsilon_{0}+P_{0}\right)\nabla\cdot u_{0}+\epsilon_{0}'\dot{T}_{0} & = & 0,\label{equation-0-u}
\end{eqnarray}
and the other three components orthogonal to $u_{0}^{\mu}$ by projecting
Eq.(\ref{equation-0}) with $\Delta_{0\mu\nu}$
\begin{eqnarray}
\Delta_{0\mu\nu}\left(\dot{u}_{0}^{\nu}+\frac{1}{T_{0}}\partial^{\nu}T_{0}\right) & = & 0\label{equation-0-x}
\end{eqnarray}
where $\dot T_0\equiv u_0^\mu \partial_\mu u_0,\ \epsilon'_{0}\equiv\frac{d\epsilon}{dT}\Huge|_{T=T_{0}}=\frac{d\epsilon_{0}}{dT_{0}}$
and $\Delta_{0\mu\nu}\equiv g_{\mu\nu}+u_{0\mu}u_{0\nu}$. The zero-order
entropy current is given by
\begin{eqnarray}
S_{0}^{\mu} & = & \frac{P_{0}}{T_{0}}u_{0}^{\mu}-\frac{1}{T_{0}}u_{0\nu}T_{0}^{\nu\mu}=\frac{\epsilon_{0}+P_{0}}{T_{0}}u_{0}^{\mu}=s_{0}u_{0}^{\mu}.\label{entropy-1}
\end{eqnarray}
It is well known that it is conserved for the ideal fluid
\begin{eqnarray}
\partial_{\mu}S_{0}^{\mu} & = & 0.\label{div-S-1}
\end{eqnarray}

\subsection{The first-order equations}

Now let us continue to deal with the next-{to}-leading
order, in which the energy-momentum tensor is given by
\begin{eqnarray}
T_{1}^{\nu\mu} & = & \left(\epsilon_{0}'+P_{0}'\right)T_{1}u_{0}^{\mu}u_{0}^{\nu}+P_{0}'T_{1}g^{\mu\nu}+\left(\epsilon_{0}+P_{0}\right)\left(\bar{u}_{1}^{\mu}u_{0}^{\nu}+u_{0}^{\mu}\bar{u}_{1}^{\nu}\right)-\eta_{0}\sigma_{0}^{\mu\nu}\label{Tmunu-1}
\end{eqnarray}
where $\bar{u}_{1}^{\mu}\equiv\Delta_{0}^{\mu\nu}u_{1\nu}$ and
\begin{eqnarray}
\sigma_{0}^{\mu\nu} & \equiv & \Delta^{\mu\alpha}\Delta^{\nu\beta}\left(\partial_{\alpha}u_{0\beta}+\partial_{\beta}u_{0\alpha}-\frac{2}{3}\Delta_{\alpha\beta}\nabla\cdot u_{0}\right).
\end{eqnarray}
{Note that since there are corrections to the temperature
$T_{1}(x)$, the energy density and pressure will also have some corrections,
\begin{equation}
\epsilon(T)=\epsilon_{0}+\epsilon_{0}^{\prime}T_{1}+O(K^{2}),\; P(T)=P_{0}+P_{0}^{\prime}T_{1}+O(K^{2}).
\end{equation}
}It should be noted that we have constrained the normalization condition
for $u_{1}^{\mu}$ as
\begin{eqnarray}
 &  & \left(u_{0}+u_{1}\right)^{2}=-1,
\end{eqnarray}
which leads to the following relation:
\begin{eqnarray}
u_{0}\cdot u_{1}=1-\sqrt{1+\bar{u}_{1}^{2}}.
\end{eqnarray}
Hence only three components of $u_{1}^{\mu}$ are independent and
the component parallel to $u_{0}^{\mu}$ can be totally determined
by $\bar{u}_{1}^{\mu}$. In addition, we can notice that $u_{0}\cdot u_{1}$
only contributes to at least second order, that is why only $\bar{u}_{1}^{\mu}$
is involved in the first-order energy-momentum tensor (\ref{Tmunu-1}).

Since the zero-order energy-momentum tensor has already {satisfied}
the conservation equation, we need the first-order energy-momentum
tensor to satisfy the conservation equation independently
\begin{eqnarray}
\partial_{\nu}T_{1}^{\nu\mu} & = & 0.\label{equation-1}
\end{eqnarray}
The component parallel to $u_{0}^{\mu}$ reads
\begin{eqnarray}
 &  & \nabla\cdot\bar{u}_{1}+\frac{1}{T_{0}}\left(\frac{1}{v_{s}^{2}}+1\right)\bar{u}_{1}^{\mu}\partial_{\mu}T_{0}
 +\frac{1}{T_{0}v_{s}^{2}}\dot{T}_{1}+{T_{1}}\left(\frac{1}{T_{0}v_{s}^{2}}\right)'\dot{T}_{0}=\frac{1}{\epsilon_{0}+P_{0}}u_{0\mu}C_{1}^{\mu},
 \label{equation-1-u}
\end{eqnarray}
where
\begin{eqnarray}
u_{0\mu}C_{1}^{\mu}=\eta_{0}\sigma_{0}^{\mu\nu}\partial_{\nu}u_{0\mu}
\end{eqnarray}
and the components orthogonal to $u_{0}^{\mu}$ reads
\begin{eqnarray}
 &  & \Delta_{0\mu\alpha}\left(\dot{\bar{u}}_{1}^{\alpha}+\bar{u}_{1}^{\nu}\partial_{\nu}u_{0}^{\alpha}\right)
 +\frac{1}{T_{0}}\bar{u}_{1\mu}\dot{T}_{0}+\frac{1}{T_{0}}\Delta_{0\mu\alpha}\partial^{\alpha}T_{1}
 -\frac{T_{1}}{T_{0}^{2}}\Delta_{0\mu\nu}\partial^{\nu}T_{0}=\frac{1}{\epsilon_{0}+P_{0}}\Delta_{0\mu\alpha}C_{1}^{\alpha}
 \label{equation-1-x}
\end{eqnarray}
where
\begin{eqnarray}
\Delta_{0\mu\alpha}C_{1}^{\alpha}=\Delta_{0\mu\alpha}\partial_{\nu}\left(\eta_{0}\sigma_{0}^{\nu\alpha}\right).
\end{eqnarray}
The first-order correction to the entropy current $S_{0}^{\mu}$ is
given by
\begin{eqnarray}
S_{1}^{\mu} & = & \frac{1}{T_{0}}\left[\epsilon_{0}'T_{1}u_{0}^{\mu}+\left(\epsilon_{0}+P_{0}\right)\bar{u}_{1}^{\mu}\right].\label{entropy-1}
\end{eqnarray}
It is easy to show that the divergence of the entropy current is always
positive and consistent with the second thermal law,
\begin{eqnarray}
\partial_{\mu}S_{1}^{\mu} & = & \frac{\eta_{0}}{2T_{0}}\sigma_{0}^{\mu\nu}\sigma_{0\mu\nu}\geq0.\label{div-S-1}
\end{eqnarray}

\subsection{The second-order equations}

We now turn to the second order, in which the energy-momentum tensor
is given by
\begin{eqnarray}
T_{2}^{\nu\mu} & = & \left(\epsilon_{0}'+P_{0}'\right)T_{2}u_{0}^{\mu}u_{0}^{\nu}+P_{0}'T_{2}g^{\mu\nu}+\left(\epsilon_{0}+P_{0}\right)\left(\bar{u}_{2}^{\mu}u_{0}^{\nu}+u_{0}^{\mu}\bar{u}_{2}^{\nu}\right)\nonumber \\
 &  & +\frac{1}{2}T_{1}^{2}\left[P_{0}''g^{\mu\nu}+(\epsilon_{0}''+P_{0}'')u_{0}^{\mu}u_{0}^{\nu}\right]+\left(\epsilon_{0}+P_{0}\right)\left(\bar{u}_{1}^{2}u_{0}^{\mu}u_{0}^{\nu}+\bar{u}_{1}^{\mu}\bar{u}_{1}^{\nu}\right)\nonumber \\
 &  & +\left(\epsilon_{0}'+P_{0}'\right)T_{1}\left(\bar{u}_{1}^{\mu}u_{0}^{\nu}+\bar{u}_{0}^{\mu}u_{1}^{\nu}\right)-\eta_{0}'T_{1}\sigma_{0}^{\mu\nu}-\eta_{0}\sigma_{1}^{\mu\nu}+\Pi_{0}^{\nu\mu},
\end{eqnarray}
where $\bar{u}_{2}^{\mu}\equiv\Delta_{0}^{\mu\nu}u_{2\nu}$. {For
energy density and pressure, we have $\epsilon=\epsilon_{0}+\epsilon_{0}^{\prime}T_{1}+\epsilon_{0}^{\prime}T_{2}+\frac{1}{2}\epsilon_{0}^{\prime\prime}T_{1}^{2}$,
and $P=P_{0}+P_{0}^{\prime}T_{1}+P_{0}^{\prime}T_{2}+\frac{1}{2}P_{0}^{\prime\prime}T_{1}^{2}$.}
Just as we did for $u_{1}^{\mu}$, we have constrained the normalization
condition for $u_{2}^{\mu}$ as
\begin{eqnarray}
\left(u_{0}+u_{1}+u_{2}\right)^{2}=-1,
\end{eqnarray}
which results in
\begin{eqnarray}
u_{0}\cdot u_{2}=\sqrt{1+\bar{u}_{1}^{2}}-\sqrt{1+\bar{u}_{1}^{2}+2\bar{u}_{1}\cdot\bar{u}_{2}+\bar{u}_{2}^{2}}.
\end{eqnarray}
It is easy to show that $u_{0}\cdot u_{2}$ only contributes to at
least third order, which can be dropped off for the second order $T_{2}^{\mu\nu}$.
However, we must consider $u_{0}\cdot u_{1}$ which has been neglected
at the first order $T_{1}^{\mu\nu}$. Since both the zero-order and
the first order energy-momentum tensors have already satisfy the
conservation equation, we need the second-order energy-momentum tensor to
satisfy the conservation equation independently, i.e.,
\begin{eqnarray}
\partial_{\nu}T_{2}^{\nu\mu} & = & 0.\label{equation-2}
\end{eqnarray}
The component parallel to $u_{0}^{\mu}$ reads,
\begin{eqnarray}
 &  & \nabla\cdot\bar{u}_{2}+\frac{1}{T_{0}}\left(\frac{1}{v_{s}^{2}}+1\right)\bar{u}_{2}^{\mu}\partial_{\mu}T_{0}
 +\frac{1}{T_{0}v_{s}^{2}}\dot{T}_{2}+{T_{2}}\left(\frac{1}{T_{0}v_{s}^{2}}\right)'\dot{T}_{0}=\frac{1}{\epsilon_{0}+P_{0}}u_{0\mu}C_{2}^{\mu},
 \label{equation-2-u}
\end{eqnarray}
where
\begin{eqnarray}
u_{0\mu}C_{2}^{\mu} & = & \epsilon_{0}u_{0}^{\mu}\partial_{\mu}\left[\frac{\left(\epsilon_{0}+P_{0}\right)}{2}\left(\frac{1}{\epsilon_{0}+P_{0}}\right)''T_{1}^{2}-\bar{u}_{1}^{2}\right]\nonumber \\
 &  & +u_{0}^{\mu}\partial_{\mu}\left[\frac{\left(\epsilon_{0}+P_{0}\right)}{2}\left(\frac{P_{0}}{\epsilon_{0}+P_{0}}\right)''T_{1}^{2}-P_{0}\bar{u}_{1}^{2}\right]\nonumber \\
 &  & -\left[\epsilon_{0}'T_{1}u_{0}^{\mu}+(\epsilon_{0}+P_{0})\bar{u}_{1}^{\mu}\frac{}{}\right]\partial_{\mu}\left[\frac{\left(\epsilon_{0}'+P_{0}'\right)}{\left(\epsilon_{0}+P_{0}\right)}T_{1}\right]\nonumber \\
 &  & -\left[\left(\frac{\epsilon_{0}'+P_{0}'}{\epsilon_{0}+P_{0}}\eta_{0}-\eta_{0}'\right)T_{1}\sigma_{0}^{\mu\nu}-\eta_{0}\sigma_{1}^{\mu\nu}+\pi_{0}^{\nu\mu}\right]\frac{1}{2}\sigma_{0\mu\nu}\nonumber \\
 &  & +\left(\epsilon_{0}+P_{0}\right)\bar{u}_{1}^{\nu}u_{0\mu}\partial_{\nu}\bar{u}_{1}^{\mu},
\end{eqnarray}
and the components orthogonal to $u_{0}^{\mu}$ reads
\begin{eqnarray}
 &  & \Delta_{0\mu\alpha}\left(\dot{\bar{u}}_{2}^{\alpha}+\bar{u}_{2}^{\nu}\partial_{\nu}u_{0}^{\alpha}\right)
 +\frac{1}{T_{0}}\bar{u}_{2\mu}\dot{T}_{0}+\frac{1}{T_{0}}\Delta_{0\mu\alpha}\partial^{\alpha}T_{2}
 -\frac{T_{2}}{T_{0}^{2}}\Delta_{0\mu\nu}\partial^{\nu}T_{0}=\frac{1}{\epsilon_{0}+P_{0}}\Delta_{0\mu\alpha}C_{2}^{\alpha},
 \label{equation-2-x}
\end{eqnarray}
where
\begin{eqnarray}
\Delta_{0\mu\alpha}C_{2}^{\alpha} & = & P_{0}\Delta_{0\mu\nu}\partial^{\nu}\left[-\frac{\left(\epsilon_{0}+P_{0}\right)}{2}\left(\frac{1}{\epsilon_{0}+P_{0}}\right)''T_{1}^{2}+\bar{u}_{1}^{2}\right]\nonumber \\
 &  & +\Delta_{0\mu\nu}\partial^{\nu}\left[\frac{\left(\epsilon_{0}+P_{0}\right)}{2}\left(\frac{P_{0}}{\epsilon_{0}+P_{0}}\right)''T_{1}^{2}-P_{0}\bar{u}_{1}^{2}\right]\nonumber \\
 &  & -\Delta_{0\mu\alpha}\partial_{\nu}\left[\left(\eta_{0}'-\frac{\epsilon_{0}'+P_{0}'}{\epsilon_{0}+P_{0}}\eta_{0}\right)\sigma_{0}^{\nu\alpha}+\eta_{0}\sigma_{1}^{\nu\alpha}-\pi_{0}^{\nu\alpha}\right]\nonumber \\
 &  & +T_{1}^{\nu\alpha}\Delta_{0\mu\alpha}\partial_{\nu}\left[\frac{\left(\epsilon_{0}'+P_{0}'\right)}{\left(\epsilon_{0}+P_{0}\right)}T_{1}\right]+\Delta_{0\mu\alpha}\partial_{\nu}\left[\left(\epsilon_{0}+P_{0}\right)\bar{u}_{1}^{\alpha}\bar{u}_{1}^{\nu}\right].
\end{eqnarray}
The second-order correction to the entropy current is
\begin{eqnarray}
S_{2}^{\mu} & = & -\frac{1}{T_{0}}u_{0\nu}T_{2}^{\nu\mu}+\frac{1}{2}\bar{u}_{1}^{2}S_{0}^{\mu}-\frac{T_{1}}{T_{0}}S_{1}^{\mu}
-\frac{1}{T_{0}}\bar{u}_{1\nu}T_{1}^{\nu\mu}+\frac{T_{1}}{T_{0}^{2}}\left(\frac{1}{2}\epsilon_{0}'T_{1}u_{0}^{\mu}+P_{0}'T_{0}\bar{u}_{1}^{\mu}\right).\label{entropy-1}
\end{eqnarray}
It is straightforward to derive the rate of entropy production for the second
order as
\begin{eqnarray}
\partial_{\mu}S_{2}^{\mu} & = & \frac{1}{2T_{0}}\left(\eta_{0}'T_{1}\sigma_{0}^{\mu\nu}+2\eta_{0}\sigma_{1}^{\mu\nu}-\pi_{0}^{\mu\nu}\right)\sigma_{0\mu\nu}.\label{div-S-2}
\end{eqnarray}
Generally, they are not positive definite, however they do not violate
the second law of thermodynamics since the third order terms must
be small compared to the second order term in the domain of applicability
of hydrodynamics. {Similar possible negative signs
and comments can also be found in \cite{Baier:2007ix}.}

Actually, such a recursion process can be generalized to any higher
orders without any {difficulty}.
It is important to note that all the equations have a similar form,
i.e., the component parallel to $u_{0}^{\mu}$ reads
\begin{eqnarray}
\label{eq-0-n}
& &\nabla\cdot\bar{u}_{n}+\frac{1}{T_{0}}\left(\frac{1}{v_{s}^{2}}+1\right)\bar{u}_{n}^{\mu}\partial_{\mu}T_{0}
+\frac{1}{T_{0}v_{s}^{2}}\dot{T}_{n}+{T_{n}}\left(\frac{1}{T_{0}v_{s}^{2}}\right)'\dot{T}_{0}=\frac{1}{\epsilon_{0}+P_{0}}u_{0\mu}{C}_{n}^{\mu},\label{equation-1-u}
\end{eqnarray}
and the components orthogonal to $u_{0}^{\mu}$ reads
\begin{eqnarray}
\label{eq-x-n}
 &  & \Delta_{0\mu\alpha}\left(\dot{\bar{u}}_{n}^{\alpha}+\bar{u}_{n}^{\nu}\partial_{\nu}u_{0}^{\alpha}\right)+\frac{1}{T_{0}}\bar{u}_{n\mu}\dot{T}_{0}+\frac{1}{T_{0}}\Delta_{0\mu\alpha}\partial^{\alpha}T_{n}
 -\frac{T_{n}}{T_{0}^{2}}\Delta_{0\mu\nu}\partial^{\nu}T_{0}=\frac{1}{\epsilon_{0}+P_{0}}\Delta_{0\mu\alpha}{C}_{n}^{\alpha},\label{equation-1-x}
\end{eqnarray}
where ${C}_{n}^{\alpha}$ depends only on the $u_{0}^{\mu}$, $T_{0}$,
$\bar{u}_{m}^{\mu}$, and $T_{m}$ $1\leq m\leq n-1$ or their derivatives.
It should be pointed out that our process is very similar to the method
used in \cite{Bhattacharyya:2008jc,Banerjee:2008th,Floerchinger:2014hua}.

\section{Initial conditions and stability}

In order to solve the hydrodynamic equations, we must give some specific
initial conditions, e.g., $u(t_{0},\vec{x})$ and $T(t_{0},\vec{x})$,
where $t_{0}$ is the initial time.  Generally, we can decompose them into
\begin{eqnarray}
u^{\mu}(t_{0},\vec{x}) & = & u_{0}^{\mu}(t_{0},\vec{x})+u_{1}^{\mu}(t_{0},\vec{x})+u_{2}^{\mu}(t_{0},\vec{x})+...,\nonumber \\
T(t_{0},\vec{x}) & = & T_{0}(t_{0},\vec{x})+T_{1}(t_{0},\vec{x})+T_{2}(t_{0},\vec{x})+...,
\end{eqnarray}
in any way as long as they satisfy
\begin{eqnarray}
 &  & u_{0}^{\mu}(t_{0},\vec{x})\gg u_{1}^{\mu}(t_{0},\vec{x})\gg u_{2}^{\mu}(t_{0},\vec{x})\gg...,\label{eq:decomposion-u-0}\\
 &  & T_{0}(t_{0},\vec{x})\gg T_{1}(t_{0},\vec{x})\gg T_{2}(t_{0},\vec{x})\gg...,
\end{eqnarray}
With different decompositions, the final result should differ only in higher
orders. For simplicity, we can just set
\begin{eqnarray}
\label{simple}
u_{0}^{\mu}(t_{0},\vec{x}) & = & u^{\mu}(t_{0},\vec{x}),\ \ \ u_{1}^{\mu}(t_{0},\vec{x})=0,\ \ \ u_{2}^{\mu}(t_{0},\vec{x})=0,\ \ \ ...\nonumber \\
T_{0}(t_{0},\vec{x}) & = & T(t_{0},\vec{x}),\ \ \ \ T_{1}(t_{0},\vec{x})=0,\ \ \ T_{2}(t_{0},\vec{x})=0,\ \ \ ...\label{decomposion-t-0}
\end{eqnarray}
With the initial state $u_{0}^{\mu}(t_{0},\vec{x})$, we can solve the zeroth-order equations (\ref{equation-0-u}) and (\ref{equation-0-x}) and obtain
the solution $u_{0}^{\mu}(t,\vec{x})$. With  this zeroth-order solution, we can calculate the first-order inhomogeneous term $u_{0\mu}C_{1}^{\mu}$ and $\Delta_{0\mu\alpha}C_{1}^{\alpha}$
which includes the first derivative of $u_{0}^{\mu}(t,\vec{x})$ and solve the first-order equations (\ref{equation-1-u}) and (\ref{equation-1-x}) under the initial conditions $\ u_{1}^{\mu}(t_{0},\vec{x})=0$ and $T_{1}(t_{0},\vec{x})=0$.
After getting the first-order solution, we can proceed further to obtain the second-order contribution and so on. Hence,  to solve the $n$th-order equations, there is no need to
know the initial value of the derivative of $n$th-order correction; we only  need the derivative of lower order corrections which have been solved already. This  should be a good advantage in our iterative method  compared to other methods.
Besides, using the  initial conditions (\ref{decomposion-t-0}), we actually rule out all the free modes
which will lead to  instability from the homogenous solutions in Eqs.(\ref{equation-1-u}) and (\ref{equation-1-x})
for $n\geq1$. Only the particular solution which is proportional to the inhomogeneous term survives. However, in the numerical simulation, the computation
error can  be inevitable and make the above argument invalid. The interesting thing in our method is that whether the instability arises or not
depends only on the zeroth-order solution $u_{0}^{\mu}$, as shown in Eq.(\ref{eq-0-n}) and Eq.(\ref{eq-x-n}) In the next section, we will use Bjorken
flow as a simple example to illustrate how the
 perturbations evolve.
\section{Bjorken Flow}

In this section, we will choose the (1+1)-dimensional Bjorken flow \cite{Bjorken:1982qr} as an example to illustrate the validity of our formalism.
In order to do that, we will use  the coordinate system, such that
\begin{eqnarray}
\tau=\sqrt{t^2-z^2},\ \ \ \eta=\tanh^{-1} \frac{z}{t}=\frac{1}{2}\ln\frac{t+z}{t-z}.
\end{eqnarray}
Bjorken flow is given by
\begin{eqnarray}
\label{Bjorken-u}
u^\mu(\tau)=\left(\cosh \eta,0,0,\sinh\eta\right).
\end{eqnarray}
In the following, we will  explicitly solve for the velocity field $u^\mu$ and the energy density $\epsilon$ or
temperature $T$ with the initial conditions
\begin{eqnarray}
u^\mu(\tau_0)=\left(\cosh \eta,0,0,\sinh\eta\right),\ \ \ \epsilon(\tau_0)=3P(\tau_0)=a T^4(\tau_0)=\frac{C}{\tau_0^{4/3}},
\end{eqnarray}
where $C$ is a constant and  we have used  $\epsilon_0=3P_0=a T_0^4$ with $a$ a constant for the conformal fluid.
We will choose the decomposition in Eq.(\ref{simple}), i.e.,
\begin{eqnarray}
u_{0}^{\mu}(\tau_0) & = & u^\mu(\tau_0) ,\ \ \ u_{1}^{\mu}(\tau_0)=0,
\ \ \ u_{2}^{\mu}(\tau_0)=0,\ \ \ ... ,\nonumber \\
T_{0}(\tau_0) & = & T(\tau_0),\ \ \ \ T_{1}(\tau_0)=0,\ \ \ T_{2}(\tau_0)=0,\ \ \ ...
\end{eqnarray}
Hence we have designated full initial configuration to the  the zeroth-order equations or ideal hydrodynamic equations
(\ref{equation-0-u}) and (\ref{equation-0-x}). From the uniqueness of the solution for the differential equations, the solution
must be the  Bjorken's solution
\begin{eqnarray}
\label{Bjorken-e}
u^\mu_0(\tau)=\left(\cosh \eta,0,0,\sinh\eta\right),\ \ \ \epsilon_0(\tau)=3P_0(\tau)=a T_0^4(\tau)= \frac{C}{\tau^{4/3}}.
\end{eqnarray}
Substituting the Bjorken  solutions (\ref{Bjorken-e}) into the first-order equations
(\ref{equation-1-u}) and (\ref{equation-1-x}) yields
\begin{eqnarray}
\label{equation-1-u-B}
\nabla\cdot\bar{u}_{1}+\frac{3}{T_{0}}\partial_\tau {T}_{1}+\frac{1}{\tau T_{0}}T_1
&=&\frac{\eta_0}{\epsilon_{0}\tau^2},\\
\label{equation-1-x-B}
\partial_\tau {\bar{u}}_{1\mu}+\frac{ v_{0\mu}}{\tau} ( v_{0}\cdot \bar{u}_{1})
-\frac{1}{3\tau}\bar{u}_{1\mu} + \frac{1}{T_{0}}\Delta_{0\mu\alpha}\partial^{\alpha}T_{1}
&=&0,
\end{eqnarray}
where $v_0^\mu=\left(\frac{z}{\tau},0,0,\frac{t}{\tau}\right)$.
Given the initial condition  $T_1(\tau_0)=0$ and $\bar u_1^\mu(\tau_0)=(0,0,0,0)$, we can solve the equations and obtain
\begin{eqnarray}
\label{Bjorken-1}
\bar u_1^\mu=(0,0,0,0),\ \ \
T_1 = \frac{\hat \eta_0 }{2\tau_0^{2/3}}\left[1-\left(\frac{\tau_0}{\tau}\right)^{2/3}\right]T_0,
\end{eqnarray}
where $\hat \eta_0$ and $\kappa$ are both constants and defined as in Ref.\cite{Baier:2007ix}
\begin{eqnarray}
\eta_0=b T_0^3 = C \hat \eta_0 \left(\frac{\epsilon_0}{C}\right)^{3/4}=\frac{C\hat \eta_0}{\tau},\ \ \ \kappa=\left(\frac{C}{a}\right)^{1/4}.
\end{eqnarray}
The energy density of the first order can be  given by
\begin{eqnarray}
\epsilon_1&=&\frac{2\hat \eta_0 }{\tau_0^{2/3}}\left[1-\left(\frac{\tau_0}{\tau}\right)^{2/3}\right]\epsilon_0.
\end{eqnarray}
Now substituting the first-order solution (\ref{Bjorken-1}) into the second-order equations (\ref{equation-2-u}) and (\ref{equation-2-x}),
we can have
\begin{eqnarray}
\label{equation-2-u-B-1}
\nabla\cdot\bar{u}_{2}+\frac{3}{T_{0}}\partial_\tau {T}_{2}+\frac{1}{\tau T_{0}}T_2
&=&\frac{2\left(\hat\eta_0 \hat\tau_{\Pi0}-\hat\lambda_{1,0}\right)}{3\tau^{7/3}},\\
\label{equation-2-x-B-1}
\partial_\tau {\bar{u}}_{2\mu}+\frac{ v_{0\mu}}{\tau} ( v_{0}\cdot \bar{u}_{2})
-\frac{1}{3\tau}\bar{u}_{2\mu} + \frac{1}{T_{0}}\Delta_{0\mu\alpha}\partial^{\alpha}T_{2}
&=&0,
\end{eqnarray}
where
\begin{eqnarray}
\tau_{\Pi0}=\hat \tau_{\Pi0}\left(\frac{\epsilon_0}{C}\right)^{-1/4},\ \ \ \lambda_{1,0}=C \hat \lambda_{1,0}\left(\frac{\epsilon_0}{C}\right)^{1/2}.
\end{eqnarray}
Again with the initial condition $T_2(\tau_0)=0$ and $\bar u_2^\mu(\tau_0)=(0,0,0,0)$, the solution is given by
\begin{eqnarray}
\bar u_2^\mu=(0,0,0,0),\ \ \
T_2 =\frac{\left(\hat\eta_0 \hat\tau_{\Pi0}-\hat\lambda_{1,0}\right)}{6\tau_0^{4/3}}\left[1-\left(\frac{\tau_0}{\tau}\right)^{4/3}\right]T_0,
\end{eqnarray}
or the energy density
\begin{eqnarray}
\epsilon_2&=&\frac{3\hat \eta_0^2}{2\tau_0^{4/3}}\left[1-\left(\frac{\tau_0}{\tau}\right)^{2/3}\right]^2\epsilon_0
+\frac{2\left(\hat\eta_0 \hat\tau_{\Pi0}-\hat\lambda_{1,0}\right)}{3\tau_0^{4/3}}\left[1-\left(\frac{\tau_0}{\tau}\right)^{4/3}\right]\epsilon_0.
\end{eqnarray}
Up to the second-order contribution, the energy density is given by
\begin{eqnarray}
\label{epsilon}
\epsilon&=&\frac{C}{\tau^{4/3}}\left\{1+\frac{\hat \eta_0 }{2\tau_0^{2/3}}\left[1-\left(\frac{\tau_0}{\tau}\right)^{2/3}\right]
+\frac{3\hat \eta_0^2}{2\tau_0^{4/3}}\left[1-\left(\frac{\tau_0}{\tau}\right)^{2/3}\right]^2\right.\nonumber\\
& &\left.+\frac{2\left(\hat\eta_0 \hat\tau_{\Pi0}-\hat\lambda_{1,0}\right)}{3\tau_0^{4/3}}\left[1-\left(\frac{\tau_0}{\tau}\right)^{4/3}\right]\right\}.
\end{eqnarray}
It is obvious that our expansion method will be valid as long as
\begin{eqnarray}
\frac{\hat \eta_0 }{\tau_0^{2/3}}\ll 1,\ \ \
\frac{\left|\hat\eta_0 \hat\tau_{\Pi0}-\hat\lambda_{1,0}\right|}{\tau_0^{4/3}}\ll 1.
\end{eqnarray}
Our result in Eq.(\ref{epsilon}) is  consistent with the result obtained  in Ref.\cite{Baier:2007ix} once we drop the terms including $\tau_0$ which enters
due to the constraint of the initial conditions.

Now let us take into account the stability problems in this specific example.  We will follow the method given by Gubser and Yarom in Ref.\cite{Gubser:2010ui}. In order to do that, we rewrite the same homogeneous differential equations
corresponding to Eqs.(\ref{equation-1-u-B}) and (\ref{equation-1-x-B}) or Eqs.(\ref{equation-2-u-B-1}) and (\ref{equation-2-x-B-1}) or even higher orders as

\begin{eqnarray}
\label{equation-tau-1}
\partial_\tau {\delta\hat T}+\frac{1}{3\tau}\partial_\eta \delta u_\eta +\frac{1}{3} \nabla_\perp \cdot \delta{u}_\perp &=& 0,\\
\label{equation-eta-1}
\partial_\tau {\delta{u}}_{\eta}+\frac{2}{3\tau}  \delta{u}_\eta +\frac{1}{\tau} \partial_\eta\delta \hat T &=& 0,\\
\label{equation-perp-1}
\partial_\tau {\delta{\bf u}}_{\perp}-\frac{1}{3\tau}\delta {\bf u}_{\perp}+{\bm\nabla}_{\perp}\delta \hat  T &=& 0,
\end{eqnarray}
where  $\delta\hat T=\delta T/T_0$, $\delta u_\eta = v_0\cdot \delta u$ and $\delta {\bf u}_{\perp}=(0,\delta u_x , \delta u_y, 0)$. Here we have used
$\delta T$ and $\delta u_\mu$ to denote $T_1,T_2, ..., T_n $ and $\bar u_{1\mu}, \bar u_{2\mu}, ...,\bar u_{n\mu}$ respectively.
In the momentum space,
\begin{eqnarray}
\delta\hat T=\int \delta\hat{\mathcal{T}}e^{ik_\eta \eta +i{\bf k}_\perp\cdot {\bf x}_\perp}d  k_\eta d^2{\bf k}_\perp,\ \ \
\delta\hat u_\mu=\int \delta\hat{{U}}_\mu e^{ik_\eta \eta +i{\bf k}_\perp\cdot{\bf x}_\perp}d k_\eta d^2 {\bf k}_\perp.
\end{eqnarray}
It follows that
\begin{eqnarray}
\label{equation-tau-2}
\partial_\tau \delta \hat{ \mathcal{T}}+\frac{i}{3\tau} k_\eta \delta{ {U}}_\eta +\frac{i}{3}{\bf k}_\perp \cdot \delta {\bf {U}}_\perp&=&0,\\
\label{equation-eta-2}
\partial_\tau {\delta{{U}}}_{\eta}+\frac{2}{3\tau}  \delta{{U}}_\eta +\frac{i}{\tau} k_\eta\delta \hat {\mathcal{T}} &=& 0,\\
\label{equation-perp-2}
\partial_\tau {\delta{{\bf U}}}_{\perp\mu}-\frac{1}{3\tau}\delta {\bf{U}}_{\perp\mu} +i {\bf k}_{\perp \mu}\delta \hat {\mathcal{T}}&=&0.
\end{eqnarray}
We can decompose $ {\delta{\bf{U}}}_{\perp\mu}$ into
\begin{eqnarray}
{\delta{\bf{U}}}_{\perp}={\bf k}_{\perp}\delta{ {W}} +\tilde{\bf k}_{\perp}\delta\tilde { {W}},
\end{eqnarray}
where $\tilde{\bf k}_{\perp}$ is a constant transverse vector satisfying  ${\bf k}_{\perp}\cdot \tilde {\bf k}_{\perp} =0$.
Then we find ${\delta\tilde{{ W}}}$ decouples with the other functions
\begin{eqnarray}
\label{equation-tau-2}
\partial_\tau \delta \hat{ \mathcal{T}}+\frac{i}{3\tau} k_\eta \delta{ {U}}_\eta +\frac{i}{3}{k}_\perp^2 \delta {W}&=&0,\\
\label{equation-eta-2}
\partial_\tau {\delta{{U}}}_{\eta}+\frac{2}{3\tau}  \delta{{U}}_\eta +\frac{i}{\tau} k_\eta\delta \hat {\mathcal{T}} &=& 0,\\
\label{equation-perp-3}
\partial_\tau {\delta{W}} +i \delta \hat {\mathcal T}-\frac{1}{3\tau}\delta {W} &=& 0,\\
\partial_\tau {\delta\tilde{W}} -\frac{1}{3\tau}\delta \tilde{W}&=& 0.
\end{eqnarray}
The solution for $\delta\tilde{W}$ is given by
\begin{eqnarray}
{\delta\tilde{W}}={\delta\tilde{W}}_{0}\left(\frac{\tau}{\tau_0}\right)^{1/3}.
\end{eqnarray}
We cannot  get the analytic  solutions for the other functions with the arbitrary $k_\eta $ and $k_\perp$.
However, we can take two limits $k_\eta=0$ and $k_\perp=0$.
When $k_\eta=0$, we can have
\begin{eqnarray}
{\delta{W}}&=&C_1 \tau^{1/3}J_{\frac{2}{3}}\left(\frac{k_\perp \tau }{3}\right)+C_2 \tau^{1/3}N_{\frac{2}{3}}\left(\frac{k_\perp \tau }{3}\right),\\
{\delta{U}}_{\eta}&=& C_3 \frac{1}{\tau^{2/3}},\ \ \ \delta \hat {\mathcal T}=i\left(\partial_\tau -\frac{1}{3\tau}\right)\delta W,
\end{eqnarray}
where $J_{\frac{2}{3}}$ and $N_{\frac{2}{3}}$ denote Bessel  and Neumann functions respectively and $C_1, C_2$, and $C_3$ are all integration constants.
When $k_\perp=0$, we can have the solution
\begin{eqnarray}
\delta U_\eta &=& C_4 \left(\frac{1}{\tau}\right)^{\frac{1+\sqrt{1-3k_\eta^2} }{3}}
+C_5 \left(\frac{1}{\tau}\right)^{\frac{1-\sqrt{1-3k_\eta^2} }{3}},\\
\delta \hat {\mathcal T}&=&\frac{i\tau}{k_\eta}\left(\partial_\tau + \frac{2}{3\tau}\right)\delta U_\eta,\\
{\delta{W}}&=& C_6 \tau^{1/3} +\frac{i C_4}{k_\eta} \tau^{\frac{2-\sqrt{1-3k_\eta^2} }{3}} +\frac{i C_5}{k_\eta} \tau^{\frac{2+\sqrt{1-3k_\eta^2} }{3}},
\end{eqnarray}
where $C_4, C_5$, and $C_6$ are also integration constants.
From the results of both  limits, we can noticed that the perturbations $\delta U_\eta$   and $\delta \hat {\mathcal T}$  always decay with the proper time $\tau$ increasing.
The perturbation ${\delta{\tilde W}}$ increases as $\tau^{1/3}$ with the proper time. The evolution of the perturbation ${\delta{ W}}$ is more complicated and depends on the
specific $k_\eta$ and $k_\perp$. However, it is obvious that there is no exponential increase and the behavior of the perturbation increase must be less than the first order of $\tau$.

\section{Conclusion}

In this paper, we have presented a perturbative procedure for solving
the viscous hydrodynamic equation order by order in the {framework}
of an effective theory.

{For simplicity, we only consider a conformal fluid
and more general cases can be straightforward to be obtained. Firstly,
we expand all hydrodynamic quantities and differential equations in the
power of the Knudsen number. Secondly, we assume the conservation equations
are satisfied order by order independently. In the leading order,
we get the solutions of an ideal fluid. By solving the differential equations
at first and second order, we find these equations have a uniform
expression with different sources. Therefore, we argued that our method
can be extended to any orders. }
We have  taken the Bjorken flow as an example and found that our method is very powerful and has good advantage to deal with
the initial condition and perturbation evolution.  It should be noticed that in our current work we  limited ourselves
to the theoretical analysis;  we  postpone  the complete  numerical analysis and manipulation  to a future study.

\begin{acknowledgments}
J.H.G. was supported in part by the Major State Basic Research Development
Program in China (Grant {No}. 2014CB845406), the National
Natural Science Foundation of China under the Grant No.~11105137, 11475104
and CCNU-QLPL Innovation Fund (QLPL2014P01). S.P was supported in
part by the NSFC under the Grant No.~11205150.

\appendix
\end{acknowledgments}

\section{Order expansion in kinetic theory \label{sec:Order-expansion-in}}

Our method is inspired by  microscopic kinetic theory. As a macroscopic
effective theory, hydrodynamic equations can be obtained from other
microscopic theories. In most of those microscopic theories, the differential
equations are expanded in terms of scaling, then are solved in each
order independently. As an example, let us consider the relativistic
kinetic theory Boltzmann equations without external fields,
\begin{equation}
\frac{df}{dt}\equiv\frac{p^{\mu}}{E_{p}}\partial_{\mu}f=\mathcal{C}[f],
\end{equation}
where $f$ is the distribution functions of particles, $p^{\mu}=(E_{p},\mathbf{p})$
is the four-momentum of particles and $\mathcal{C}[f]$ is the collision
term. We can expand $f$ and $\mathcal{C}[f]$ in a gradient expansion
way, $f=f_{0}+f_{1}+...$, $\mathcal{C}=\mathcal{C}_{0}+\mathcal{C}_{1},$
which is equivalent to expanding in powers of $K$. For simplicity,
we neglect the higher order terms in the collision term, and simply set
$\mathcal{C}=\mathcal{C}_{0}$. In this case, the current and energy-momentum
tensor in each order are given by the integration over momentum, i.e.,
$j_{n}^{\mu}=\int\frac{d^{3}p}{(2\pi)^{3}}\frac{p^{\mu}}{E_{p}}f_{n}$
and $T_{n}^{\mu\nu}=\int\frac{d^{3}p}{(2\pi)^{3}}\frac{p^{\mu}p^{\nu}}{E_{p}}f_{n}$,
where the lower index $n$ means the $n$-th order in the gradient
expansion. Taking covariant derivatives, we get,
\begin{eqnarray}
\partial_{\mu}j_{n}^{\mu} & = & \int\frac{d^{3}p}{(2\pi)^{3}}\frac{p^{\mu}}{E_{p}}\partial_{\mu}f_{n}=\int\frac{d^{3}p}{(2\pi)^{3}}\mathcal{C}[f_{n-1}]=0,\nonumber \\
\partial_{\mu}T_{n}^{\mu\nu} & = & \int\frac{d^{3}p}{(2\pi)^{3}}\frac{p^{\mu}p^{\nu}}{E_{p}}\partial_{\mu}f_{n}=\int\frac{d^{3}p}{(2\pi)^{3}}p^{\nu}\mathcal{C}[f_{n-1}]=0,
\end{eqnarray}
where in the last line, we  used the results of
the time-reversal symmetry of the collision term, which guarantees
 total energy-momentum and number conservation. That implies that in
each order the currents and energy-momentum tensor are conserved independently,
which is very similar to our method.



\begin{thebibliography}{10}
\bibitem{Heinz:2013th} U.~Heinz and R.~Snellings, 
 Annu.\ Rev.\ Nucl.\ Part.\ Sci.\ \textbf{63}, 123 (2013). 


\bibitem{Gale:2013da} C.~Gale, S.~Jeon and B.~Schenke, 
 Int.\ J.\ Mod.\ Phys.\ A \textbf{28}, 1340011 (2013). 








\bibitem{Teaney:2009qa} D.~A.~Teaney, 
 arXiv:0905.2433 {[}nucl-th{]}. 


\bibitem{Romatschke:2007mq} P.~Romatschke and U.~Romatschke, 
 Phys.\ Rev.\ Lett.\ \textbf{99}, 172301 (2007); 


\bibitem{Song:2007fn} H.~Song and U.~Heinz, 
 Phys.\ Lett.\ \textbf{B658}, 279 (2008); 
 Phys.\ Rev.\ C \textbf{77}, 064901 (2008); 
 Phys.\ Rev.\ C \textbf{78}, 024902 (2008); 


\bibitem{Dusling:2007gi} K.~Dusling and D.~Teaney, 
 Phys. Rev. C \textbf{77}, 034905 (2008). 


\bibitem{Molnar:2008xj} D.~Molnar and P.~Huovinen, 
 J.\ Phys.\ G \textbf{35}, 104125 (2008). 





\bibitem{Schenke:2010rr} B.~Schenke, S.~Jeon and C.~Gale, 
 Phys.\ Rev.\ Lett.\ \textbf{106}, 042301 (2011); 
 Phys.\ Rev.\ C \textbf{85}, 024901 (2012). 





\bibitem{Eckart:1940te} C.~Eckart, 
 Phys.\ Rev.\ \textbf{58}, 919 (1940). 





\bibitem{Landau:book} L.~D.~Landau and E.~M~Lifshitz,\textit{Fluid
Mechanics} (Pergamon, London, 1959)




\bibitem{Hiscock:1983zz} W.~A.~Hiscock and L.~Lindblom, 
 Annals Phys.\ \textbf{151}, 466 (1983). 
 W.~A.~Hiscock and L.~Lindblom, 
 Phys.\ Rev.\ D \textbf{31}, 725 (1985). 
 W.~A.~Hiscock and L.~Lindblom, 
 Phys.\ Rev.\ D \textbf{35}, 3723 (1987). 





\bibitem{Muller:1967} I.~Müller, Z.\ Phys. 198, 329(1967)




\bibitem{Israel:1979wp} W.~Israel and J.~M.~Stewart, 
 Annals Phys.\ \textbf{118} (1979) 341. 





\bibitem{Pu:2009fj}S.~Pu, T.~Koide and D.~H.~Rischke,
  Phys.\ Rev.\ D {\bf 81}, 114039 (2010)

\bibitem{Denicol:2008ha}G.~S.~Denicol, T.~Kodama, T.~Koide and P.~Mota,
J.\ Phys.\ G {\bf 35}, 115102 (2008)   [arXiv:0807.3120 [hep-ph]].

\bibitem{Pu:2010zz}S.~Pu, T.~Koide and Q.~Wang,   
AIP Conf.\ Proc.\  {\bf 1235}, 186 (2010).

\bibitem{Pu:2011vr}S.~Pu,
arXiv:1108.5828 [hep-ph].

\bibitem{Baier:2007ix} R.~Baier, P.~Romatschke, D.~T.~Son, A.~O.~Starinets
and M.~A.~Stephanov, 
 JHEP \textbf{0804}, 100 (2008) 





\bibitem{Betz:2008me} B.~Betz, D.~Henkel and D.~H.~Rischke, 
 Prog.\ Part.\ Nucl.\ Phys.\ \textbf{62}, 556 (2009) 





\bibitem{Betz:2009zz} B.~Betz, D.~Henkel and D.~H.~Rischke, 
 J.\ Phys.\ G \textbf{36} (2009) 064029. 





\bibitem{York:2008rr} M.~A.~York and G.~D.~Moore, 
 Phys.\ Rev.\ D \textbf{79}, 054011 (2009) 





\bibitem{Gao:2012ix}J.~H.~Gao, Z.~T.~Liang, S.~Pu, Q.~Wang and X.~N.~Wang,
Phys.\ Rev.\ Lett.\  {\bf 109}, 232301 (2012)

\bibitem{Chen:2012ca}J.~W.~Chen, S.~Pu, Q.~Wang and X.~N.~Wang,
Phys.\ Rev.\ Lett.\  {\bf 110} (2013) 26,  262301

\bibitem{Chen:2013dca}J.~W.~Chen, J.~H.~Gao, J.~Liu, S.~Pu and Q.~Wang,
Phys.\ Rev.\ D {\bf 88} (2013) 074003

\bibitem{Bhattacharyya:2008jc} S.~Bhattacharyya, V.~E.~Hubeny,
S.~Minwalla and M.~Rangamani, 
 JHEP \textbf{0802}, 045 (2008) 





\bibitem{Banerjee:2008th} N.~Banerjee, J.~Bhattacharya, S.~Bhattacharyya,
S.~Dutta, R.~Loganayagam and P.~Surowka, 
 JHEP \textbf{1101}, 094 (2011) 






\bibitem{Floerchinger:2014hua}
  S.~Floerchinger, U.~A.~Wiedemann, A.~Beraudo, L.~Del Zanna, G.~Inghirami and V.~Rolando,
  Nucl.\ Phys.\ A {\bf 931}, 965 (2014)

\bibitem{Bjorken:1982qr}
  J.~D.~Bjorken,
  Phys.\ Rev.\ D {\bf 27}, 140 (1983).
\bibitem{Gubser:2010ui}
  S.~S.~Gubser and A.~Yarom,
  Nucl.\ Phys.\ B {\bf 846}, 469 (2011)


\end{thebibliography}
\end{document}